\documentclass[journal]{IEEEtran}
\UseRawInputEncoding
\usepackage{blindtext}
\usepackage{cite}
\usepackage{amsmath,amssymb,amsfonts}
\usepackage[noend]{algpseudocode}
\usepackage{graphicx}
\usepackage{epstopdf}
\usepackage{textcomp}
\usepackage[linesnumbered,ruled,vlined]{algorithm2e}
\usepackage{diagbox}
\usepackage{svg}
\usepackage{xcolor}
\definecolor{my_brown}{RGB}{115,115,115}
\usepackage{makecell}
\usepackage{booktabs}     
\usepackage{amsmath}      
\usepackage{amssymb}      
\usepackage{graphicx}     
\usepackage{caption}      
\usepackage{float}        
\usepackage{array}        
\usepackage{multirow}     
\usepackage{adjustbox}    
\usepackage{longtable}    
\begin{document}

\title{Hybrid Cognitive IoT with Cooperative Caching and SWIPT-EH: A Hierarchical Reinforcement Learning Framework}

\author{Nadia~Abdolkhani,~\IEEEmembership{Student Member, ~IEEE}
        and~Walaa~Hamouda,~\IEEEmembership{Senior~Member,~IEEE}
\thanks{N. Abdolkhani, and W. Hamouda are with the Department of Electrical and Computer Engineering, Concordia University, Montreal, QC, H3G 1M8, Canada (e-mail: n\_abdolk@ece.concordia.ca; hamouda@ece.concordia.ca).}
}


\maketitle

\begin{abstract}
\textcolor{black}{This paper proposes a hierarchical deep reinforcement learning (DRL) framework based on the soft actor-critic (SAC) algorithm for hybrid underlay-overlay cognitive Internet of Things (CIoT) networks with simultaneous wireless information and power transfer (SWIPT)-energy harvesting (EH) and cooperative caching. Unlike prior hierarchical DRL approaches that focus primarily on spectrum access or power control, our work jointly optimizes EH, hybrid access coordination, power allocation, and caching in a unified framework. The joint optimization problem is formulated as a weighted-sum multi-objective task, designed to maximize throughput and cache hit ratio while simultaneously minimizing transmission delay. In the proposed model, CIoT agents jointly optimize EH and data transmission using a learnable time switching (TS) factor. They also coordinate spectrum access under hybrid overlay-underlay paradigms and make power control and cache placement decisions while considering energy, interference, and storage constraints. Specifically, in this work, cooperative caching is used to enable overlay access, while power control is used for underlay access. A novel three-level hierarchical SAC (H-SAC) agent decomposes the mixed discrete-continuous action space into modular subproblems, improving scalability and convergence over flat DRL methods. The high-level policy adjusts the TS factor, the mid-level policy manages spectrum access coordination and cache sharing, and the low-level policy decides transmit power and caching actions for both the CIoT agent and PU content. Simulation results show that the proposed hierarchical SAC approach significantly outperforms benchmark and greedy strategies. It achieves better performance in terms of average sum rate, delay, cache hit ratio, and energy efficiency, even under channel fading and uncertain conditions.}
\end{abstract}


\section{Introduction}
\IEEEPARstart{T}{he} rapid expansion of smartphones, cloud services, and the Internet of Things (IoT) has dramatically intensified the demand for wireless spectrum. To address this growing pressure, cognitive radio (CR) technology has emerged as a promising solution by allowing unlicensed secondary users (SUs) to opportunistically access underutilized licensed spectrum. When CR capabilities are integrated into IoT frameworks, they give rise to cognitive IoT (CIoT) networks, where intelligent devices dynamically manage both spectrum and energy resources while coexisting with primary users (PUs)~\cite{nada_survey_2023}.

A common characteristic of CIoT devices is that they are often energy-constrained and deployed in remote or power-limited environments. In such settings, energy harvesting (EH) becomes vital for ensuring long-term operation. EH enables CIoT devices to gather energy from ambient radio frequency (RF) signals, providing a sustainable alternative to battery replacement. Compared to traditional wireless power transfer (WPT), simultaneous wireless information and power transfer (SWIPT) allows for a more efficient use of time slots by enabling EH and data transmission within the same slot through a time-switching (TS) protocol. This dual functionality requires devices to make intelligent real-time decisions regarding TS ratios and transmission power. However, many underlay CR models in existing studies focus only on individual aspects such as access coordination, power control, or EH optimization, without considering the advantages of jointly optimizing these processes for improved performance and spectrum efficiency \cite{9680720, 10064678, chu_eh_ciot_ma_2019, 9238884}.

Overlay CR models, on the other hand, promote spectrum access through cooperation, typically by having SUs relay PU data. While this paradigm can improve spectrum utilization, it also introduces practical challenges, such as handover delays, sensitivity to PU activity, and dependence on limited backhaul infrastructure~\cite{Farooq_relaying_Access_2024, Liang_relaying_Ucom_2024, Wilson_relaying_ICCC_2024, 9893136}. To mitigate these issues, edge caching has gained attention as a complementary technique. By pre-storing frequently accessed content near end users, edge caching reduces both latency and backhaul load. However, conventional caching strategies often assume static or pre-defined popularity profiles~\cite{Li_cache_nonlearning_2021, Yang_cache_nonlearning_2019, Yang_cache_nonlearning_2020, Nissar_cache_gametheory_2019}, which are unrealistic in dynamic CIoT environments where content demand fluctuates over time and across locations.
Recognizing the limitations of using either underlay or overlay spectrum access alone, hybrid CR frameworks have emerged as a more practical and flexible solution \cite{9493626, 10744388}. These frameworks allow CIoT devices to dynamically switch between EH, cooperative relaying, and direct spectrum access based on current network conditions, PU activity, and internal resource constraints. This adaptability enables SUs to make context-aware decisions that improve throughput, reduce delay, and respect interference limits. 

\textcolor{black}{Optimization-based baselines are frequently reported in related work. They typically presume full and accurate knowledge of system dynamics (e.g., instantaneous channel states) to cast the problem as a convex or finite combinatorial program. In our setting, however, the dynamics are stochastic and the decisions are mixed discrete-continuous, making exact optimization intractable without strong, unrealistic assumptions.} Effectively managing this complex behavior requires intelligent and adaptive decision-making frameworks. Reinforcement learning (RL) offers a model-free paradigm for CIoT agents to learn optimal strategies by interacting with the environment and receiving feedback in the form of rewards. Deep reinforcement learning (DRL) methods, such as deep deterministic policy gradient (DDPG), proximal policy optimization (PPO), and multi-agent variants, have been applied to tasks such as spectrum access, power control, and cooperative caching in distributed CIoT systems~\cite{Gao_Liu_coopcaching_2022}. However, these approaches often struggle with scalability, high-dimensional action spaces, and instability, especially in real-time, resource-constrained settings.
To address these challenges, hierarchical reinforcement learning (HRL) has emerged as a more scalable and robust alternative. HRL decomposes complex decision-making tasks into multiple layers of abstraction, enabling agents to manage high-dimensional environments more effectively. One of HRL's key advantages is its ability to handle mixed action spaces, allowing for both discrete and continuous decisions. This hierarchical architecture not only enhances learning efficiency and training stability but also accelerates convergence and simplifies long-term credit assignment. As a result, HRL has gained increasing attraction in CIoT applications that demand adaptability, efficiency, and robustness in dynamic spectrum and energy management scenarios.

\subsection{Related Work}
In CIoT environments, energy sustainability is a critical concern. In~\cite{9205906}, the authors propose a cooperative SWIPT framework tailored for 6G CIoT networks while using orthogonal frequency-division multiple access (OFDMA) to access the primary network. However, RL is not employed in their solution. Several works have employed flat DRL models to address this challenge. While these studies demonstrate improvements in energy efficiency and network throughput, they are typically built on flat control architectures and limited to fixed access strategies. For instance, Tashman et. al \cite{Tashman_2023_SWIPT_constant} address the problem of optimizing performance in SWIPT-EH underlay CR networks where SUs must effectively balance opportunistic transmission with limited harvested energy. A model-free RL approach is employed to derive an optimal power control policy while assuming a constant TS factor. 
Zheng et al.~\cite{Zheng2023_Hybrid_Backscatter_EH_CRN} propose a communication scheme that combines ambient backscatter and harvest-then-transmit paradigms for SUs in EH CR networks. The authors apply a model-free Q-learning algorithm to solve it to maximize the long-term average throughput while adhering to interference constraints for PUs. 

In~\cite{Sun2020_coop_WPT_large_CIoT}, the authors investigate WPT in a large-scale cognitive industrial IoT network with cooperative spectrum sharing. They adopt a cooperative co-evolutionary algorithm, a bio-inspired optimization technique, to solve the resource allocation problem aimed at maximizing throughput and energy efficiency (EE) under underlay spectrum access. Another related study~\cite{Chen2023_UAV_WPT_DDPG} presents a modern DDPG-based DRL framework to optimize UAV-assisted WPT in CR-enabled IoT networks, jointly targeting throughput and energy efficiency.
More recent studies, such as~\cite{10564176}, explore hybrid-action DRL techniques under imperfect channel state information, emphasizing the need for robust learning models to handle real-world uncertainties. A more integrated approach is presented in~\cite{Nadia2025_SWIPT_underlay}, which addresses joint optimization of time and power allocation in a SWIPT-enabled CIoT network with EH capabilities. A double deep Q-Network (DDQN) enhanced with an upper confidence bound (UCB) exploration strategy is employed to balance energy consumption and data transmission while abiding by the constraints. 
 
The work in ~\cite{9013975} explores DRL-based content delivery in cache-enabled heterogeneous networks (HetNets), though the study is not specific to CR networks.
The work in~\cite{8540871} addresses the problem of SWIPT for CIoT networks. It proposes a resource allocation framework that jointly optimizes time and power splitting ratios to maximize data rate and energy efficiency. The approach is based on mathematical optimization and does not employ any learning algorithms. Similarly,~\cite{Liu2019_EE_Coop_CRN} investigates energy-efficient task offloading in mobile edge computing (MEC)-enabled CR networks, where secondary users relay primary user data in exchange for spectrum access. This method is formulated using convex optimization and does not incorporate SWIPT or learning-based techniques.
Shukla et al.\cite{Shukla2022_SWIPT_CIoT_Relay} focus on EE communication in a SWIPT-enabled CIoT networks employing non-orthogonal multiple access (NOMA). They introduce a coordinated direct and relay transmission scheme to improve throughput and reliability through optimization-based resource allocation, without using RL. In\cite{9765586}, the authors perform an analytical outage probability analysis of a SWIPT-enabled two-way cooperative CR network using relay techniques, also without involving learning methods. Further,~\cite{Nadia_ICC_overlay_coopcache} addresses the challenge of enhancing service availability and reducing latency in overlay CIoT networks by allowing SUs to cooperate in caching content requested by PUs, rather than simply vacating the spectrum. The proposed framework introduces a DRL algorithm based on DDQN with a Zipf adjusted UCB exploration (DDQN-UCBZ), enabling efficient trade-offs between exploration and exploitation.

\textcolor{black}{Hierarchical DRL has recently been explored extensively for IoT and wireless networks owing to its modularity and faster convergence in large decision spaces. For instance,~\cite{9824994} introduces a hierarchical federated learning (FL) framework for user-edge association, where DRL agents manage the trade-off between energy consumption and learning accuracy. Similarly,~\cite{Nadia_GC_HRL} develops a hierarchical DDPG algorithm that decomposes the action space into channel selection, WPT-EH mode, and power control to cope with jamming attacks.
Security aspects are addressed in~\cite{10678783}, which models interactions among legitimate users, eavesdroppers, and third-party collaborators through a hierarchical game with DRL-based matching and coalition formation. In MEC-assisted hierarchical FL,~\cite{9764370} proposes a reward-structured DDPG algorithm, coupled with NOMA, to jointly minimize latency, energy, and model inaccuracy. Hierarchical UAV control is studied in~\cite{10923644}, where a two-tier framework employs wulti-agent DQN for sensor selection and advantage actor-critic (A2C) for trajectory optimization to reduce the Age of Information. Cooperative edge caching is considered in~\cite{9447004}, which leverages hierarchical clustering and federated DRL to balance latency reduction and content diversity. Likewise,~\cite{HRL_CNN_DQN_IoT} applies a hierarchical DQN enhanced by convolutional neural networks (CNN) to the deployment of the service function chain, achieving energy and carbon efficiency in large-scale IoT. Finally, spectrum allocation in integrated Terrestrial Networks (TNs) and Non-Terrestrial Networks (NTN) systems is tackled in~\cite{11016260}, where global, regional, and local agents coordinate through PPO to improve throughput, fairness, and spectral efficiency.}

Hybrid CR networks, which combine underlay and overlay access models, have been investigated to enhance flexibility in spectrum utilization.  Works such as~\cite{Zheng2023_Hybrid_Backscatter_EH_CRN,Pardeep2023_Hybrid_bascatter_CRN} explore innovative hybrid interweave-underlay CR frameworks that integrate ambient backscatter communication and EH to maximize system throughput. A dynamic hybrid spectrum access protocol is introduced in~\cite{10510683}, where access decisions are made based on spectrum sensing metrics. Although this work does not employ DRL, its rule-based access logic provides motivation for hierarchical learning frameworks.
Similarly, the authors in~\cite{Sun2022_Adaptive_SWIPT_CIoT} address resource allocation in a CIoT network with SWIPT capabilities. They propose a hybrid access model that combines underlay and overlay transmission and formulate a non-convex optimization problem to maximize EE. The method is optimization-based and does not involve RL or caching. In another study,~\cite{Hlapisi2023_CIoT_Hybrid} explores throughput optimization in CIoT networks through analytical modeling of a hybrid spectrum access scheme. The framework addresses varying interference constraints but does not employ EH or RL. These works collectively highlight the value of hybrid access design, although the absence of learning-based and integrated approaches limits their adaptability in dynamic CIoT environments.

\subsection{Motivation and Contributions}
\textcolor{black}{According to the above analysis of related works, hierarchical DRL has been widely explored in IoT networks, but prior work differs from ours in scope and method. Both our study and~\cite{Nadia_GC_HRL} use three-level hierarchical DRL for an energy-constrained SU with hybrid actions. However,~\cite{Nadia_GC_HRL} adopts hierarchical DDPG with a smart jammer in an underlay setting, whereas we use hierarchical soft actor-critic (H-SAC) with entropy regularization, hybrid overlay-underlay access, and cooperative caching. While~\cite{9824994} applies hierarchical RL to energy-aware client-edge association in FL, we target CIoT spectrum access optimizing throughput, delay, and cache performance under EH and interference. \cite{10678783} models DRL-aided coalition formation for physical-layer security, while we focus on throughput, caching, and energy—not security. \cite{9764370} addresses mobile edge computing (MEC)-assisted hierarchical FL with NOMA to reduce latency and model error, whereas we tackle CIoT coexistence with cooperative caching and EH constraints. \cite{10923644} optimizes UAV trajectories to reduce age of information, but we address CIoT spectrum-energy coordination. \cite{9447004} uses hierarchical federated DRL with clustered devices for cooperative caching, while we embed caching within the agent’s own hierarchical DRL in a CIoT network. \cite{HRL_CNN_DQN_IoT} targets service function chains deployment via hierarchical CNN-DQN (network virtualization), whereas we pursue physical-layer resource control via H-SAC. Finally,~\cite{11016260} allocates spectrum across multi-tier TN/NTN agents, while our hierarchy decomposes decisions within a single CIoT agent coexisting with PUs.}

To the best of our knowledge, no prior work has explored the joint integration of caching, power control, SWIPT-based EH, and hybrid spectrum access within a unified hierarchical DRL framework for CIoT networks. The main contributions of this work are summarized as follows:
\begin{itemize}
    \item We introduce a CIoT communication architecture that operates under a hybrid CR paradigm, combining both underlay and overlay access strategies. This hybrid architecture enhances spectrum utilization, improves adaptability to dynamic wireless environments, and enables cooperative strategies. Also, we introduce caching of PU content as a bargaining mechanism to enable cooperation with the PU and gain spectrum access under overlay access paradigms. 


    \item The joint optimization problem is formulated as a weighted-sum multi-objective optimization task that simultaneously aims to maximize the throughput and cache hit ratio while minimizing transmission delay. We propose a unified model that jointly optimizes EH, access coordination, cache management, and power control in a dynamic and complex wireless environment. The framework operates under practical constraints, including interference limitations, finite cache capacity, unpredictable PU activity, and energy availability challenges.


    \item We develop a novel three-level H-SAC framework to mitigate the action-space explosion commonly encountered in flat DRL architectures, and to effectively handle mixed-type actions (discrete and continuous). This hierarchical design enhances convergence stability, reducing computational and energy overhead, and significantly increases the probability of discovering optimal policies compared to conventional flat DRL methods. 

\end{itemize}
The rest of this paper is structured as follows. Section II outlines the system model. The problem formulation is
detailed in Section III, followed by the proposed H-SAC algorithm in Section IV. Section V presents the performance
evaluation of our framework, and conclusions are drawn in Section VI.

\section{System Model}
\subsection{Cognitive IoT Network}
\begin{figure}
    \centering
    \includegraphics[width=1\linewidth]{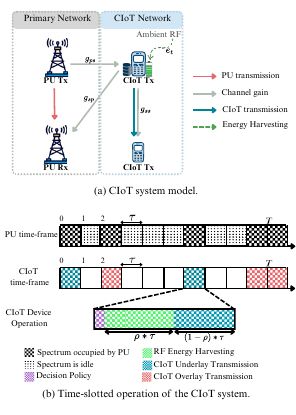}
    \caption{System Model}
    \label{fig:sys_mod}
\end{figure}

Consider the time-slotted CIoT network depicted in Fig.~\ref{fig:sys_mod}, which consists of a transmitter-receiver (Tx-Rx) pair representing the CIoT agent, coexisting with a PU Tx-Rx pair. Following the time structure in \cite{Nadia_IoT_jamm_2024}, the communication timeline is divided into $T$ equal-duration time slots indexed by $t = 1, 2, \ldots, T$, with each slot lasting \(\tau\) seconds. The PU Tx is allocated $L$ time slots for its transmissions, during which it operates at a fixed transmit power \(P^t_p\). The PU's activity during each time slot $t$ is indicated by a binary variable \(\omega_p^t\), where \(\omega_p^t = 1\) if the PU is active, and \(\omega_p^t = 0\) otherwise. The PU occupancy pattern over the time horizon is represented by the binary vector \(\mathbf{w}_{\text{p}} \in \{0, 1\}^{T}\).

The wireless channels are modeled as independent and identically distributed (i.i.d.) Rayleigh fading. \textcolor{black}{This model is well-suited for CIoT scenarios characterized by dense deployments, low transmit powers, small/omnidirectional antennas, and predominantly non-line-of-sight (NLoS) links.} Specifically, the channel power gains for the links between the CIoT Tx and Rx \(g_{ss}^t\), the PU Tx and CIoT Rx \(g_{ps}^t\), the CIoT Tx and PU Rx \(g_{sp}^t\) follow exponential distributions. The probability density function (PDF) of the gain \(g_{ij}^t\) is given by
\[
f_{g_{ij}^t}(y) = h_{ij} \exp(-h_{ij} y),
\]
where \(h_{ij} = d_{ij}^{-\alpha}\) is the channel fading parameter determined by the distance \(d_{ij}\) between nodes \(i\) and \(j\), and the path-loss exponent \(\alpha\). The channel is assumed to remain constant during each slot \cite{9645987}.

The CIoT Tx is equipped with a finite battery of size \(B_{\max}\) (Watt) and supports SWIPT through the TS protocol. In each time slot \(t\), a TS factor \(\rho_t \in [0, 1]\) is selected to divide the slot into two phases: an EH phase of duration \(\rho_t \tau\), and a data transmission phase of duration \((1 - \rho_t)\tau\). This dynamic allocation enables the CIoT agent to balance EH and communication according to its current state and environment conditions.

\subsection{Transmission Model}
We consider a hybrid CR access model where the CIoT agent can transmit data either via cooperative overlay access or underlay spectrum sharing, depending on the PU's activity and local caching status.

\paragraph*{1) Idle Channel Access}
When the PU channel is idle during time slot \(t\), and the CIoT agent's requested content is cached locally, the CIoT agent accesses the entire channel bandwidth \(W\). The achievable data rate is
\begin{equation}
\label{eq:r0_overlay}
R_0^t = \log_2\left(1 + \frac{P_s^t g_{ss}^t}{\sigma^2}\right),
\end{equation}
where \(P_s^t\) is the CIoT transmit power and \(\sigma^2\) is the noise variance. If the caching condition is not met, the CIoT agent must offload its request from the core network, which is assumed to incur higher latency and is beyond the scope of this work.

\paragraph*{2) Cooperative Overlay Access}
If the PU is active and the CIoT agent has cached both its own and the PU’s content, the PU permits cooperative access to a fraction of the bandwidth \cite{Nadia_ICC_overlay_coopcache}. In this case, the achievable rate becomes
\begin{equation}
\label{eq:r2_overlay}
R_1^t = \frac{1}{k} \log_2\left(1 + \frac{P_s^t g_{ss}^t}{\sigma^2}\right),
\end{equation}
where \(k \geq 1\) reflects the extent of bandwidth division between the PU and CIoT agent. If neither caching condition is met, the CIoT agent must offload its request from the core network.

\paragraph*{3) Underlay Spectrum Access} 
When the PU is active and the CIoT agent has cached only its own content, the agent may employ underlay spectrum access. In this mode, the CIoT agent is allowed to transmit concurrently with the PU, as long as the interference caused at the PU receiver remains below a predefined threshold \(I_{\text{th}}\). This constraint is expressed as
\begin{equation}
\label{eq:interference_constraint}
P_s^t g_{sp}^t \leq I_{\text{th}}.
\end{equation}
Under these interference conditions, the CIoT receiver also experiences interference from the active PU. Consequently, the achievable rate for the CIoT transmission is reduced and is given by
\begin{equation}
\label{eq:r1_underlay}
R_2^t = \log_2\left(1 + \frac{P_s^t g_{ss}^t}{P_p^t g_{ps}^t + \sigma^2}\right).
\end{equation}
This model enables the CIoT agent to dynamically select its access mode based on channel occupancy, caching availability, and interference constraints.

\subsection{Caching Model}
The PU and CIoT networks operate independently, with users interested in distinct content libraries. Let \(\mathcal{M} = \{1, 2, \ldots, M\}\) and \(\mathcal{N} = \{1, 2, \ldots, N\}\) denote the sets of primary and secondary content files, respectively. \textcolor{black}{For simplicity, we assume all content files are of uniform size. However, this assumption can be relaxed by dividing larger files into equal-sized chunks \cite{Nadia_ICC_overlay_coopcache, 9893136}. This assumption simplifies analysis while enabling generalization of the model and ensuring consistency with Zipf-based content popularity distributions.}
The content popularity distributions for PU and CIoT libraries are denoted by \(\Delta_p\) and \(\Delta_s\), respectively, and are initially unknown to the agent. Content request patterns for PUs and CIoT devices denoted by the binary vectors \(D_p^t = \{d_{p1}^t, d_{p2}^t, \ldots, d_{pM}^t\}\) and \(D_s^t = \{d_{s1}^t, d_{s2}^t, \ldots, d_{sN}^t\}\) in time slot \(t\) follow a Zipf distribution with skew parameters \(\zeta_p\) and \(\zeta_s\), and arrival rates \(\lambda_p\) and \(\lambda_s\) (in requests per slot), respectively \cite{Yang_cache_nonlearning_2019}. 

Let \(C_s\) represent the cache capacity of the CIoT agent. At each time slot \(t\), the CIoT agent applies a cooperative caching policy represented by two binary vectors, \(c_p^t = \{c_{p1}^t, c_{p2}^t, \ldots, c_{pM}^t\}\) caching policy for PU content and
    \(c_s^t = \{c_{s1}^t, c_{s2}^t, \ldots, c_{sN}^t\}\): caching policy for CIoT content.
Each element in the caching vectors is defined as
\[
c_{pm}^t = 
\begin{cases}
1, & \text{if the } m\text{-th PU file is cached at time slot } t, \\
0, & \text{otherwise},
\end{cases}\]

\[{c}_{sn}^t = 
\begin{cases}
1, & \text{if the } n\text{-th CIoT file is cached at time slot } t, \\
0, & \text{otherwise}.
\end{cases}
\]

The CIoT agent updates its cache content dynamically across time slots to maximize content availability while adhering to the cache capacity constraint, \( \sum_{m=1}^{M} c_{pm}^t+I^t\sum_{n=1}^{N} c_{sn}^t \leq C_s\), where $I^t$ is the cache sharing decision. The caching decision plays a vital role in enabling cooperative spectrum access and reducing reliance on backhaul communication. To quantify the performance of caching decisions, we define two key metrics: the \textit{PU Cooperation Rate (PCR)} representing the ability of the CIoT agent to serve PU requests locally and the \textit{CIoT Hit Rate (CHR)} for CIoT requests.

\paragraph*{1) PU Cooperation Rate (PCR)}
This metric measures how often the CIoT agent is able to fulfill PU content requests locally, which enables cooperative access to the licensed spectrum. The PCR over the time horizon \(T\) is then given by
\begin{equation}
\label{eq:pcr}
\text{PCR} = \frac{1}{L} \sum_{t=1}^{T} \omega_p^t c_p^t D_p^t.
\end{equation}

\paragraph*{2) CIoT Hit Rate (CHR)}
This metric measures the fraction of CIoT content requests that are satisfied directly from the local cache without requiring retrieval from the core network. The average cache hit rate over a time horizon \(T\) is defined as
\begin{equation}
\label{eq:chr}
\text{CHR} = \frac{1}{T} \sum_{t=1}^{T} c_s^t D_s^t.
\end{equation}

The aforementioned metrics are crucial for evaluating the system's ability to reduce backhaul load, enhance latency performance, and enable cooperative spectrum access. A high CHR implies better CIoT user experience, while a high PCR increases the opportunity for spectrum reuse through collaboration with the PU.

The \textit{cache offloading latency} experienced by the CIoT is calculated as \(l_s^t = \frac{u}{R^t}\),
where \( u \) denotes the content size and \( R^t \) is the achieved transmission rate at time slot \( t \).
To evaluate the latency reduction provided by local caching, we compare \( l_s^t \) with the latency of retrieving the same content from the base station (BS) and core network, which is \(l_{\text{BS}}^t = \frac{u}{R_{\text{BS}}^t}\),
where \( R_{\text{BS}}^t \) represents the backhaul transmission rate from the BS.
Our goal is to minimize the total latency by maximizing local throughput \( R^t \), thereby reducing \( l_s^t \) and avoiding the higher latency \( l_{\text{BS}}^t \) associated with BS offloading \cite{8629363}. Thus, the delay reduction is given by

\begin{align}
\mathcal{L}^t = l_{BS}^t - l_s^t.
\end{align}
Furthermore, the overall throughput is given by
\begin{align}
R^t =\ & c^t_s D^t_s (1 - \rho_t) \tau \Big[ (1 - \omega_{p}^t) R_0^t \nonumber +\ \omega_{p}^t R_1^t c^t_p D^t_p + \\
& \omega_{p}^t R_2^t (1 - c^t_p D^t_p) \Big].
\end{align}
\subsection{Energy Harvesting Model}
The CIoT Tx is equipped with a finite-capacity battery denoted by \(B_{\max}\), which is replenished through RF EH. At the start of operation, the harvested energy is assumed to be zero \(e_0=0\). Similar to \cite{Nadia_IoT_jamm_2024}, the ambient energy source is modeled as \(\hat{e} \sim U(0, E_{max})\), where $E_{max}$ is the maximum possible energy to be harvested. The actual amount of energy harvested during time slot \(t\) is given by
\[
e_t = \mu \hat{e},
\]
where \(0 \leq \mu \leq 1\) denotes the energy conversion efficiency factor. \textcolor{black}{We adopt a linear EH model, a widely used abstraction in the wireless EH and SWIPT literature [2], [41]. It captures the proportional relationship between incident RF power and harvested energy while keeping the analysis feasible and the benchmarking fair \cite{9562262}. We acknowledge that practical EH circuits are nonlinear but the extension lies beyond the scope of this paper.} The harvested energy in each slot is stored in the battery and is reserved exclusively for future operations. As noted in \cite{9606870}, this process does not consume additional energy from the PUs or nearby devices.

Let \(B_0\) denote the initial battery level, and \(B_t\) represent the energy available in the battery at the beginning of time slot \(t\). Following the modeling approach in \cite{chu_eh_ciot_ma_2019}, we assume the battery to be ideal, i.e., there are no storage or retrieval losses. The CIoT Tx consumes energy solely for data transmission. Any harvested energy that exceeds the battery capacity is discarded. Moreover, we adopt normalized time slots, allowing us to treat power and energy interchangeably.

Accordingly, the battery update equation is given by
\begin{equation}
B_{t+1} = \min\left\{
    B_t + \rho_t e_t \tau - (1 - \rho_t) P_s^t \tau,\ B_{\max}
\right\},
\end{equation}
where \(P_s^t\) denotes the transmission power selected, and \(\rho_t\) is the TS factor selected during time slot \(t\). To ensure energy feasibility, the CIoT Tx must satisfy the constraint
\begin{equation}
0 \leq (1 - \rho_t) P_s^t \tau \leq B_t.
\end{equation}
Finally, the cumulative energy consumption over the entire horizon of \(T\) time slots must not exceed the total harvested energy plus the initial battery level
\begin{equation}
\sum_{t=1}^{T} (1 - \rho_t) P_s^t \tau \leq B_0 + \sum_{t=0}^{T - 1} \rho_t e_t \tau.
\end{equation}

To evaluate the trade-off between energy consumption and performance, we define the EE of the CIoT transmitter over a horizon of \(T\) time slots as
\begin{equation}
\mathrm{EE} = \frac{\sum_{t=1}^{T} R^t}{\sum_{t=1}^{T} (1 - \rho_t) P_s^t \tau},
\end{equation}
where \(R_t\) represents the instantaneous data rate (in bits/second/Hz) achieved in time slot \(t\). 
This metric captures how efficiently the system converts energy into useful throughput. A higher EE indicates better energy utilization, which is particularly important for battery-constrained CIoT devices relying on ambient energy sources. The goal of the proposed framework is not only to maximize throughput but also to maintain sustainable operation by optimizing this efficiency under varying PU activity and environment dynamics. 

\section{Problem Formulation: A Weighted-Sum Multi-Objective Problem}

In this section, we formulate a multi-objective optimization problem to develop a dynamic strategy to maximize the throughput, cache hit rate, the cooperation rate, and reduce the latency. We consider a hierarchical decision-making problem for a CIoT Tx operating in a hybrid cooperative CR network. The CIoT agent aims to maximize a weighted sum of performance objectives, namely throughput, cooperation (PU service), cache hit rate, and EE, under a set of realistic constraints, including interference, caching capacity, and battery limitations. The overall multi-objective problem is formulated as
\begin{subequations}
\label{eq:weighted_optim}
\begin{align}
    &\max_{\substack{\rho_t, P_s^t, \\ c_{pm}^t, c_{sn}^t, I^t}} 
    \sum_{t=1}^{T} \Big\{ 
    w_1 \cdot R^t 
    + w_2 \cdot (\text{PCR}^t+\text{CHR}^t) 
    + w_3 \cdot \mathcal{L}^t \Big\} \label{eq:objective}\\
    &\text{s.t. } 
    \sum_{t=1}^{T} (1-\rho_t)\tau P_s^t \leq B_0 + \sum_{t=0}^{T-1} \rho_t e_t \tau, \label{eq:energy_budget} \\
    &\quad\quad 0 \leq (1-\rho_t)\tau P_s^t \leq B_t, \quad \rho_t \in [0, 1], \label{eq:energy_slotwise} \\
    &\quad\quad (1-I^t)\omega_{p}^t g_{sp}^t P_s^t \leq I_{\text{th}}, \quad \omega_{p}^t \in \{0,1\},I^t \in \{0,1\} \label{eq:interference} \\
    &\quad\quad c_{pm}^t, D_p^t \in \mathcal{M}, \quad c_{sn}^t, D_s^t \in \mathcal{N}, \label{eq:cache_hit_check} \\
    &\quad\quad \sum_{m=1}^{M} c_{pm}^t + I^t\sum_{n=1}^{N} c_{sn}^t \leq C_s. \label{eq:cache_capacity}
\end{align}
\end{subequations}
\textcolor{black}{The optimization problem in (13) is subject to several practical constraints that capture the key physical and system-level limitations of the considered CIoT network. The (13b) battery dynamic constraint ensures that the total transmission energy over the horizon does not exceed the initial battery level plus the accumulated harvested energy. The constraint (13c) prevents the agent from allocating more energy than is physically available in the battery. The interference constraint in (13d), enforces that the interference caused at the PU receiver remains below the predefined threshold $I_{th}$ whenever underlay access is used. The cache capacity constraint (13f), reflects the practical storage limitation and ensures realistic caching decisions, where the sum of cached PU and CIoT files must not exceed the finite cache capacity $C_s$.}

In (\ref{eq:weighted_optim}), $w_1$, $w_2$, and $w_3$ are the weights assigned to the throughput, cooperation/CIoT hit rate, and delay terms, respectively. We assume a weighted-sum multi-objective problem to dynamically maximize the throughput and the cache hit rate while reducing the delay. Improving the performance of CIoT agents by leveraging various
attributes, and the relationship between them may directly affect the agent's action decisions, remains a challenging problem. This challenge motivates the use of the weighted-sum approach for solving multi-objective problems \cite{9408025}. \textcolor{black}{The weights ($w_1$,$w_2$,$w_3$) reflect system priorities among throughput, cache efficiency, and latency reduction. In this work, they are set to (0.4, 0.3, 0.3) to balance rate-centric and cache-centric performance. Increasing $w_1$ biases the solution toward higher throughput at the expense of cache hit ratio, while larger $w_2$ improves cache efficiency and cooperation with PUs, sometimes at the cost of delay.} Additionally, \(I^t\) denotes the binary decision variable indicating whether the CIoT agent shares its cache capacity with the PU. Specifically, \(I^t = 0\) implies that the CIoT does not share its cache, thereby requiring access to the spectrum via underlay mode. Conversely, \(I^t = 1\) indicates that the CIoT cooperatively shares its cache with the PU, enabling spectrum access through the overlay mode, provided that the PU's requested file is available in the CIoT cache.

\section{Proposed Method}
\subsection{The Model-Free Markov Decision Process}
The hybrid CIoT network optimization problem is formulated as a Markov decision process (MDP), defined by the tuple \((\mathcal{S}, \mathcal{A}, \mathcal{P}, \mathcal{R}, \gamma_{disc})\), where:
\(\mathcal{S}\) is the state space representing the observable environment status, including battery level, channel conditions, and PU slot occupancy.
\(\mathcal{A}\) is the action space, consisting of TS factor, access coordination, caching decisions, and power control.
\(\mathcal{P}: \mathcal{S} \times \mathcal{A} \rightarrow \mathcal{S}\) is the (unknown) state transition probability distribution.
\(\mathcal{R}: \mathcal{S} \times \mathcal{A} \rightarrow \mathbb{R}\) is the reward function capturing throughput, delay, and cache hit ratio
and \(\gamma_{disc} \in [0,1)\) is the discount factor that balances immediate and future rewards. Moreover, the CIoT network lacks knowledge of the state transition probabilities for each occupancy state of the primary network, making it nearly impossible to accurately determine \(\boldsymbol{\mathcal{P}}\). Due to the unknown and dynamic nature of the environment (e.g., time-varying channels, non-stationary PU activity, and stochastic energy arrivals), we adopt a \textit{model-free} RL approach, where agents learn optimal policies solely based on observed transitions. Thus, the model-free MDP tuple is represented as \((\boldsymbol{\mathcal{S}},\boldsymbol{\mathcal{A}},\boldsymbol{\mathcal{R}},T)\) with the following components

\begin{enumerate}
    \item \textbf{State Space $\boldsymbol{\mathcal{S}}$:}  
The state space $\mathcal{S}$ represents all possible states of the environment over $T$ time slots. At each time step $t$, the CIoT agent observes a state $s_t$ characterized by the following components:  
the current battery level $B_t$,  
the amount of energy harvested in the previous time slot $e_{t-1}$,  
the occupancy status of the PU $\omega_p^t$,  
and the channel power gains $g_{ps}^t$ (PU-to-CIoT), $g_{sp}^t$ (CIoT-to-PU), and $g_{ss}^t$ (CIoT-to-CIoT).  
Therefore, the state at time slot $t$ can be expressed as
\begin{equation}
s_t = \{ B_t, e_{t-1}, \omega_{p}^t, g_{ps}^t, g_{sp}^t, g_{ss}^t \}.
\end{equation}
    \item \textbf{Action Space $\boldsymbol{\mathcal{A}}$:}  
The action space $\mathcal{A}$ includes all control decisions available to the CIoT agent at each time slot. Based on the current state $s_t$, the agent selects a continuous TS factor $\rho_t$ to manage the trade-off between EH and data transmission. It also chooses a binary cooperation mode $I_t$, where $I_t = 1$ corresponds to cooperative overlay access involving cache sharing with the PU, and $I_t = 0$ corresponds to non-cooperative underlay access without cache sharing. Additionally, the agent determines the transmit power level $P_s^t$, which must respect interference constraints in underlay mode. The caching policy for the CIoT agent, denoted by $c_{sn}^t$, is always selected, while the PU caching policy $c_{pm}^t$ is only determined if cooperative access ($I_t = 1$) is chosen.  

Accordingly, the action taken by the CIoT agent at time slot $t$ is represented as
\begin{equation}
a_t = [\rho_t, I_t, P_s^t, c_{pm}^t, c_{sn}^t].
\end{equation}
    \item \textbf{Reward $\boldsymbol{\mathcal{R}}$:}  
The immediate reward $r_t$ is defined as a weighted combination of key performance indicators, reflecting both the efficiency of the system and adherence to operational constraints. 
\textcolor{black}{The reward function is designed to embed the objectives specified in (\ref{eq:weighted_optim}) directly into the learning process. If the agent selects actions that satisfy all feasibility constraints (13b)-(13f). Conversely, any infeasible action, such as transmitting with insufficient battery or exceeding the PU interference threshold, incurs a penalty $-\phi$. This ensures that the agent not only maximizes system performance but also learns to operate within realistic energy, interference, and storage limitations. For instance, when the battery level $B_t$ is low, the high-level agent tends to increase the TS factor $\rho_t$ to harvest more energy, even though this reduces immediate throughput. These trade-offs are directly captured in the optimization framework of (13a)-(13f).}

\item \textbf{Time Step $\boldsymbol{T}$:}  
Each transition from time slot $t$ to $t+1$ is treated as a discrete time step. The CIoT agent's interactions are evaluated sequentially across all time slots $T$, with each state-action pair examined at every step.
\end{enumerate}

\subsection{Proposed Hierarchical Soft Actor-Critic Strategy}

In cooperative caching and energy-constrained CR networks, agents must make decisions over large, structured, and interdependent action spaces involving EH, access coordination, power control, and content caching. These decisions naturally span both continuous (e.g., transmit power, energy allocation) and discrete (e.g., access mode selection) action types. Learning a single flat policy to jointly optimize over this mixed and high-dimensional space poses significant challenges, including poor credit assignment, sample inefficiency, and unstable convergence. Moreover, most DRL algorithms are specialized for either discrete or continuous action spaces, but not both simultaneously.
To address these challenges, we propose a hierarchical reinforcement learning (HRL) framework that decomposes the global decision-making task into three meaningful levels. This decomposition allows us to use continuous-action algorithms like soft actor-critic (SAC) at the high and low levels, while applying a discrete-action algorithm such as deep Q-network (DQN) at the mid level. By aligning each sub-policy with the most suitable DRL technique, the hierarchical structure enhances modularity, learning efficiency, and overall policy performance.

Our proposed hierarchical structure is as follows:

\textbf{1. High-Level Agent:}  
The high-level agent determines the TS factor $\rho_t \in [0,1]$, which specifies the portion of time slot $t$ allocated for EH, while the remaining duration $(1 - \rho_t)$ is used for data transmission. Since this is a continuous action, we model it using a stochastic policy compatible with continuous control. We adopt the SAC algorithm at this level because the decision is highly context-dependent, influenced by factors such as battery level, channel conditions, and traffic demand. Using a continuous policy enables smoother exploration and allows for gradient-based optimization through the reparameterization trick \cite{10668814}. Moreover, SAC's entropy-regularized objective \cite{10777063} encourages diverse energy allocation behaviors during training, improving robustness and long-term performance in dynamic environments. This design allows the agent to effectively learn adaptive strategies that balance EH and data transmission over time.

The high-level SAC agent optimizes the entropy-regularized return
\begin{align}
J(\pi_{\text{high}}) = \sum_{t} \mathbb{E}_{s_t \sim \mathcal{D}} \Big[ \mathbb{E}_{\rho_t \sim \pi_{\text{high}}} \big[ 
& Q(s_t, \rho_t) \nonumber \\
& - \alpha_{ee} \log \pi_{\text{high}}(\rho_t | s_t) \big] \Big],
\end{align}
where \( \rho_t \) is the high-level action sampled from the policy \( \pi_{\text{high}} \) given state \( s_t \), and \( Q(s_t, \rho_t) \) represents the expected return. The term \( \alpha_{ee} \log \pi_{\text{high}}(\rho_t | s_t) \) adds entropy regularization, encouraging the policy to remain stochastic and explore alternative decisions rather than converging prematurely to suboptimal actions. The scalar coefficient \( \alpha_{ee} \) controls the trade-off between reward exploitation and exploration. This formulation allows the high-level agent to adaptively select long-term strategies, such as TS factors, based on observed system states, enhancing robustness in dynamic CIoT environments.

The Q-function targets are as
\begin{align}
y = r_t + \gamma_{\text{disc}} \, \mathbb{E}_{\rho_{t+1} \sim \pi_{\text{high}}} \Big[ 
& \min_i Q_{\bar{\theta}_i}(s_{t+1}, \rho_{t+1}) \nonumber \\
& - \alpha_{ee} \log \pi_{\text{high}}(\rho_{t+1} | s_{t+1}) \Big].
\end{align}
Here, \( r_t \) is the immediate reward, and \( \gamma_{disc} \) is the discount factor that controls the influence of future returns. The expectation is taken over the next high-level action \( \rho_{t+1} \) sampled from the current policy. The minimum operator over target Q-networks \( Q_{\bar{\theta}_i} \) implements double Q-learning, which helps reduce overestimation bias in value estimation. The entropy term \( \alpha_{ee} \log \pi_{\text{high}}(\rho_{t+1} | s_{t+1}) \) ensures consistent encouragement of exploration in future time steps. This target value is used to minimize the temporal difference error in critic updates, enabling stable and efficient learning of long-term value estimates.

\textbf{2. Mid-Level Agent:}  
This agent selects between overlay ($I_t = 1$) and underlay ($I_t = 0$) access modes, determining whether the CIoT user will cooperate with the PU or access the spectrum independently. Since the action space is binary and discrete, Q-learning is particularly efficient and stable for this setting. Employing a SAC approach here would introduce unnecessary complexity and computational overhead without providing additional benefits. Therefore, we adopt a DQN with $\epsilon$-greedy exploration strategy to handle this decision-making process, ensuring both simplicity and fast convergence. The agent is trained by minimizing the Q-learning loss
\begin{equation}
J_{\text{mid}}(\theta) = \mathbb{E}_{(s_t, a_t, r_t, s_{t+1}) \sim \mathcal{D}} \left[ \left( Q_\theta(s_t, a_t) - y_t \right)^2 \right],
\end{equation}
where the target is:
\begin{equation}
y_t = r_t + \gamma_{disc} \max_{a'} Q_{\bar{\theta}}(s_{t+1}, a').
\end{equation}

\textbf{3. Low-Level Agent:}  
The low-level agent is responsible for determining both the transmit power level $P_s^t \in [P_{\min}, P_{\max}]$ and the caching decisions for PU and CIoT content $c_p^t, c_s^t$, thus the low-level action is defined as $a_t^{low}=[P_s^t,c_p^t,c_s^t]$. The low-level SAC agent optimizes the entropy-regularized return
\begin{align}
J(\pi_{\text{low}}) = \sum_{t} \mathbb{E}_{s_t \sim \mathcal{D}} \bigg[ 
& \mathbb{E}_{a_t^{\text{low}} \sim \pi_{\text{low}}} \big[ Q(s_t, a_t^{\text{low}}) \nonumber \\
& - \alpha_{ee} \log \pi_{\text{low}}(\rho_t \mid a_t^{\text{low}}) \big] \bigg],
\end{align}
with Q-function targets as:
\begin{align}
y =\ & r + \gamma_{disc}\ \mathbb{E}_{a^{\text{low}}_{t+1} \sim \pi_{\text{low}}} \bigg[ \min_i Q_{\bar{\theta}_i}(s_{t+1}, a^{\text{low}}_{t+1}) \nonumber \\
& \qquad\qquad -\alpha_{ee} \log \pi_{\text{low}}(a^{\text{low}}_{t+1} \mid s_{t+1}) \bigg].
\end{align}

While caching is inherently a binary decision process, each file is either stored or not, the use of discrete-action RL methods such as DQN becomes impractical due to the combinatorial explosion of possible cache configurations. As the number of cacheable files increases, the action space grows exponentially, making it difficult for discrete methods to represent and explore effectively. To overcome these limitations, we adopt a continuous relaxation approach within the SAC framework. Specifically, the agent outputs real-valued logits for each caching decision, which are passed through a sigmoid activation and then discretized using a Top-$k$ operator. This relaxation allows us to leverage gradient-based learning while ensuring feasible binary caching outputs.
Due to the continuous nature of power control and the relaxed caching formulation, the SAC algorithm is well-suited for this level. SAC provides stable off-policy learning, supports continuous actions, and incorporates entropy regularization, enabling robust exploration and improved convergence in high-dimensional spaces. The use of continuous relaxation followed by deterministic discretization has proven effective in various domains, including differentiable sorting and structured prediction, and allows reinforcement learning agents to optimize over discrete domains while retaining differentiability during training \cite{sander_topk}.

\begin{figure}
    \centering
    \includegraphics[scale=1.1]{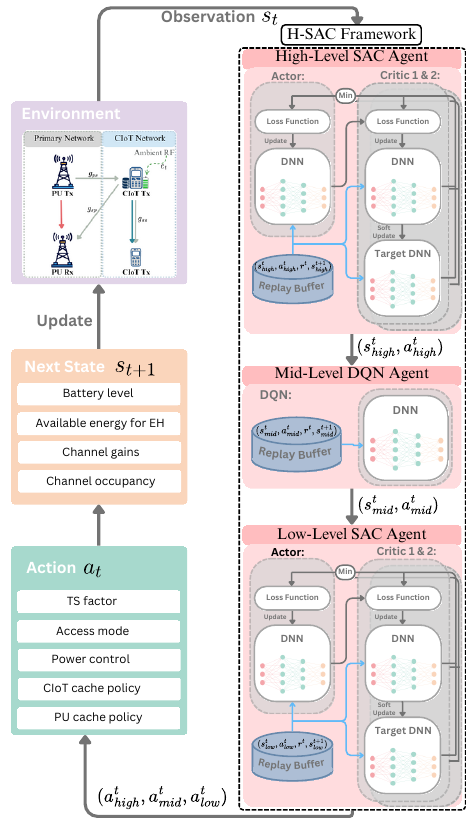}
    \caption{Proposed H-SAC Strategy}
    \label{fig:hsac}
\end{figure}

To visually and procedurally illustrate our proposed approach, we include both a schematic diagram and a detailed training algorithm. Fig.~\ref{fig:hsac} presents the hierarchical architecture of our framework, highlighting the flow of decision-making across the high-level, mid-level, and low-level agents. Complementing the figure, Algorithm.~\ref{alg:algorithm1} describes the step-by-step training procedure used to learn the hierarchical policy. 
\begin{algorithm}[t!]
\caption{Training Procedure for Our Proposed H-SAC Strategy}\label{alg:algorithm1}
\textbf{Input:} CIoT environment simulator and parameters.

\textbf{Output:} Optimal action $a_t$ in each time slot $t$.

Initialize environment and replay buffers for each agent

Initialize SAC agents: $\pi_{\text{high}}, \pi_{\text{low}}$, and DQN agent: $\pi_{\text{mid}}$

\For{episode= 1,...,episodes}{
  
    \For{t= 1,...,T}{
        Observe the state $s_t$;
        \If{buffer is not full}{
        Sample a random action $a_t$;
        }

        \Else{
        High-level action: $\rho_t \sim \pi_{\text{high}}(s_t)$ ;
         
         Mid-level state: $s_t^{\text{mid}} \gets [s_t, \rho_t]$;
         
        Mid-level action: $I^t \sim \pi_{\text{mid}}(s_t^{\text{mid}})$ ;
        
        Low-level state: $s_t^{\text{low}} \gets [s_t^{\text{mid}}, I^t]$;
        
        Low-level action: $(P_s^t, c_p^t, c_s^t) \sim \pi_{\text{low}}(s_t^{\text{low}})$;}

            Compute weighted reward; 
            
            Observe the next state $s_{t+1}$ ;
        
            Store transitions in replay buffers;
            
             Update each agent:
             
         \quad $\pi_{\text{high}}$ via SAC update;
         
         \quad $\pi_{\text{mid}}$ via DQN update;
         
         \quad $\pi_{\text{low}}$ via SAC update;}

         Log averaged metrics: reward, throughput, cache hits, delay, energy efficiency;
}
\end{algorithm}
\subsection{Complexity Analysis}
The total complexity scales with the number of environment time steps and the per-time-step computation. Let $d_s$ be the state dimension and $d_a^{\ell}$ the action dimension at level $\ell\in\{\text{high},\text{mid},\text{low}\}$. Each network uses $L$ hidden layers with $h$ neurons. For a fully connected MLP, a single forward pass with input size $d_{\text{in}}$ and output size $d_{\text{out}}$ costs
\[
\mathcal{O}\!\big(d_{\text{in}}h + (L-1)h^2 + h\,d_{\text{out}}\big).
\]
Each SAC agent has one actor and two critics. The actor input is $d_s$ and output is $d_a^{\ell}$:
\[
C_{\text{act}}^{\ell}=\mathcal{O}\!\big(d_s h + (L-1)h^2 + h\,d_a^{\ell}\big).
\]
Each critic takes $(s,a)$ as input (size $d_s+d_a^{\ell}$) and outputs a scalar:
\[
C_{\text{crit}}^{\ell}=\mathcal{O}\!\big((d_s{+}d_a^{\ell})h + (L-1)h^2 + h\big).
\]
The mid-level uses DQN, with one Q-network of output size $2$, consisting of no actor/target-actor terms with cost
\[
\mathcal{O}\big(d_s h + (L-1)h^2 + 2h\big).
\]

In terms of memory usage, in RL implementations, the two main sources of memory consumption are replay buffers and network parameters. A single transition stores $(s,a,r,s',d)$ of size $\Theta(d_s + d_a^{\ell} + d_s + 2)=\Theta(2d_s + d_a^{\ell})$. 

\textcolor{black}{Compared to flat SAC or DQN methods, the hierarchical design reduces the effective action dimension at each level. This decomposition improves sample efficiency and accelerates convergence, as shown in Section V. The added overhead comes from maintaining multiple networks (high-, mid-, and low-level agents) and replay buffers, but this is offset by faster learning and improved long-term stability. Thus, the complexity performance trade-off favors H-SAC in large, mixed-action CIoT scenarios.}

\section{Simulation Model and Results}
To assess the performance of the proposed H-SAC strategy, we conduct simulations within an EH-enabled CIoT network environment described in Section~II. The simulation framework incorporates realistic parameters such as channel noise, path loss, device distances, and interference thresholds, as well as system-level parameters including battery capacity, EH limits, and transmission settings. Additionally, caching characteristics, content request patterns, and learning configurations for all three decision-making agents, high-level SAC, mid-level DQN, and low-level SAC, are specified to reflect practical constraints and diverse action spaces. All simulation and agent configuration details are summarized in Table~\ref{tab:simulation-params}. Most parameters were adopted from commonly used values in the literature \cite{9680720,Nadia2025_SWIPT_underlay}. Unless otherwise specified, the parameter values remain consistent across all strategies in the reported results as shown in Table~\ref{tab:simulation-params}.

To assess the effectiveness of the proposed H-SAC framework for optimizing our proposed hybrid EH-enabled CIoT network, we conduct a comparative evaluation against several baseline strategies: \\
1) Hierarchical Twin Delayed Deep Deterministic Policy Gradient (H-TD3): In this variant, the SAC component in the high-level and low-level agents are replaced with TD3.\\
2) \textcolor{black}{Hierarchical Deep Deterministic Policy Gradient (H-DDPG): In this variant, the SAC component in the high-level and low-level agents are replaced with DDPG \cite{Nadia_GC_HRL}.}\\
3) 2-layer H-SAC strategy: In this variant, we assume that the time-switching factor $\rho_t$ is fixed at $0.5$ and not learnable, following the approach in \cite{Tashman_2023_SWIPT_constant}, while keeping the rest of the proposed H-SAC framework unchanged. Consequently, this strategy excludes the high-level agent and relies only on the mid-level and low-level agents, which are defined in the same manner as in the proposed H-SAC strategy.\\
4) Random Strategy: This strategy entails the CIoT agent selecting an action $a_t$ at each time step randomly from the action space, without any cognitive or intelligent decision-making involved.

To evaluate the performance of the proposed framework, we consider several key metrics: average sum rate (ASR), average delay, CHR, PCR, and EE. These metrics are computed using a weighted moving average scheme to track their progression throughout training. The moving average is updated according to
\begin{equation}
    \text{average}_{\text{new}} = (1 - \delta) \times \text{average}_{\text{old}} + \delta \times \text{value},
\end{equation}
where \( \delta \) is a smoothing factor that emphasizes recent data, while \( 1 - \delta \) retains the influence of past observations. In our setup, we use \( \delta = 0.01 \). This formulation dampens the impact of short-term variations and highlights long-term trends in system performance. 
\begin{table}[ht]
\centering
\caption{Simulation Parameters and Agent Configurations}
\label{tab:simulation-params}
\begin{tabular}{l|l}
\hline
\textbf{Parameter} & \textbf{Value} \\
\hline
Channel noise variance ($\sigma^2$) & $10^{-3}$ \\
Path-loss exponent ($\alpha$) & $4$ \\
Device distance $d_{sp}$ & $1.8\,\text{m}$ \\
Device distance $d_{ps}$ & $1.8\,\text{m}$ \\
Device distance $d_{ss}$ & $1.5\,\text{m}$ \\
Number of time slots ($T$) & $30$ \\
Time slot duration ($\tau$) & $1\,\text{s}$ \\
PU occupied slots ($L$) & $18$ \\
PU transmit power ($P_p$) & $0.2\,\text{W}$ \\
Interference threshold ($I_{\text{th}}$) & $0.01\,\text{W}$ \\
Initial harvested energy ($e_0$) & $0$ \\
Initial battery level ($B_{0}$) & $0$ \\
Battery capacity ($B_{\max}$) & $0.5\,\text{W}$ \\
Maximum harvested energy ($E_{\max}$) & $0.1\,\text{W}$ \\
Time-switching ratio ($\rho_t$) & $\{0, 0.1, \dots, 1\}$ \\
Transmit power ($P_s^t$) & $\{0, 0.01, \dots, 0.1\}$ \\
Energy efficiency factor ($\mu$) & $0.9$ \\
Number of files ($M$) & $30$ \\
Number of requests ($N$) & $30$ \\
Cache capacity ($C_s$) & $20$ \\
PU Zipf skew parameter ($\zeta_p$) & $0.8$ \\
CIoT Zipf skew parameter ($\zeta_s$) & $0.6$ \\
PU Request arrival rate ($\lambda_p$) & $15$ \\
CIoT Request arrival rate ($\lambda_s$) & $15$ \\
Bandwidth division factor ($k$) & $2$ \\
Number of episodes & $2000$ \\
Penalty for constraint violation ($\phi$) & $1$ \\
Replay buffer size& $10000$ \\
Discount factor ($\gamma_{disc}$) & $0.99$ \\
Hidden size & $256$ \\
Learning rate & $3 \times 10^{-4}$ \\
Batch size & $256$ \\
Smoothing factor ($\delta$) & $0.01$\\
Content size ($u$) & $1$\\
Multi-objective weights ($w_1,w_2,w_3$) & $0.4,0.3,0.3$\\
Exploration-exploitation coefficient ($\alpha_{ee}$) & $0.2$\\
\hline
\end{tabular}
\end{table}

\begin{figure}
    \centering
    \includegraphics[width=1\linewidth]{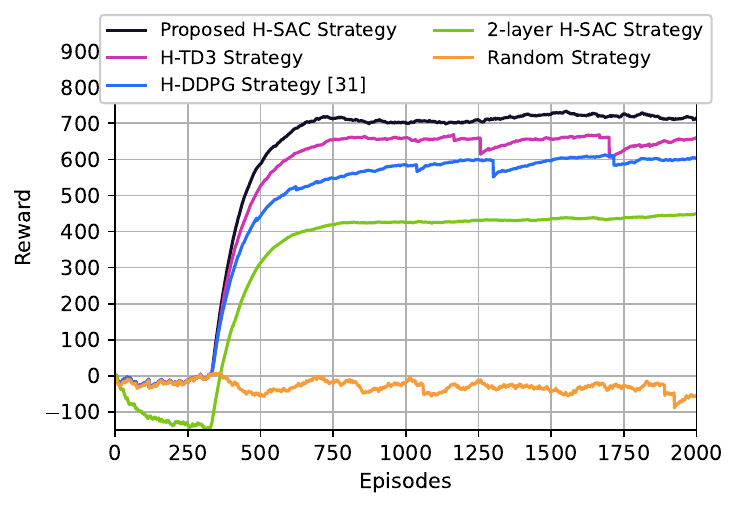}
    \caption{Benchmarking the achievable reward of our proposed 
    H-SAC strategy in comparison to the existing strategies.}
    \label{fig:Benchmarking01}
\end{figure}

Fig.~\ref{fig:Benchmarking01} presents the learning curves of different strategies in terms of cumulative reward per episode across training episodes. The proposed H-SAC strategy consistently outperforms all baselines, achieving higher reward at convergence, confirming the effectiveness of the proposed H-SAC framework compared to existing baseline strategies. For all strategies, the agent selects random actions to populate the replay buffer initially. Once the buffer is sufficiently filled, the learning process begins, which occurs at approximately episode 333. The H-TD3 and H-DDPG methods exhibit performance improvements over 2-layer H-SAC and random baselines but converge to lower reward values compared to H-SAC. H-TD3 shows better stability than H-DDPG, likely due to its target policy smoothing and delayed updates, which mitigate the overestimation bias common in actor-critic methods. 
In contrast, the 2-layer H-SAC strategy maintains a flat and relatively low reward, highlighting the limitations of non-adaptive, static policy for \(\rho\) in dynamic environments. This highlights that even if only one metric is considered constant instead of learning, the performance is considerably affected. Furthermore, the 2-layer H-SAC strategy initially yields negative rewards during the early episodes, as the selected actions incur penalties before effective learning begins. This can be attributed to the fixed TS factor assumption, which forces the agent to transmit even with an empty battery, whereas other strategies have the flexibility to harvest more energy before transmitting in the initial episodes. Similarly, the random strategy fails to improve over time and remains the lowest performer.

\begin{figure}
    \centering
    \includegraphics[width=1\linewidth]{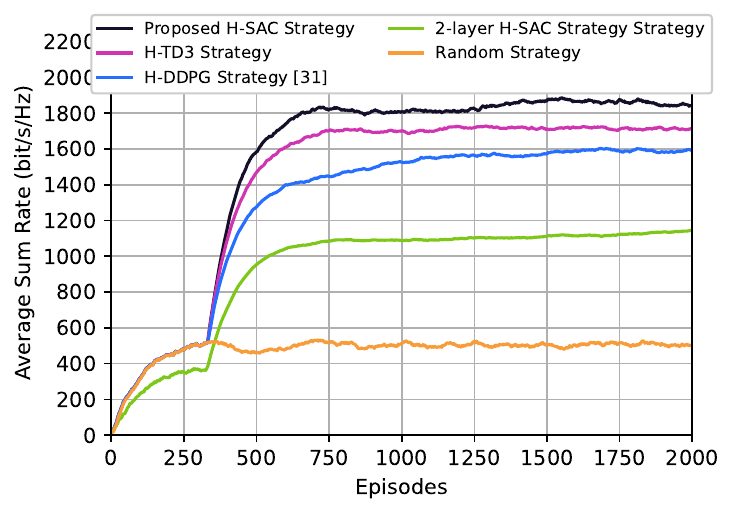}
    \caption{Benchmarking the ASR performance of our proposed 
    H-SAC strategy in comparison to other existing strategies.}
    \label{fig:Benchmarking02}
\end{figure}

Fig.~\ref{fig:Benchmarking02} illustrates the ASR performance across training episodes for the proposed H-SAC framework and several benchmark strategies. As seen in the plot, the H-SAC strategy achieves the highest ASR. This highlights its ability to learn efficient policies for jointly managing EH, power control, caching policies, and hybrid access coordination. This reflects the framework’s capability to exploit favorable conditions while minimizing interference and resource conflicts, resulting in superior spectrum efficiency.
The H-TD3 and H-DDPG strategies achieve moderate improvements in throughput but fall short of H-SAC’s performance. While H-TD3 yields a higher sum rate than H-DDPG, both methods lack the entropy regularization of H-SAC, which limits their long-term spectral efficiency.
The 2-layer H-SAC strategy, which maintains a constant TS factor of $0.5$, performs significantly worse, achieving a consistently low ASR. This highlights the limitations of static, non-adaptive strategies in dynamic CIoT environments. The random strategy remains the lowest performer throughout training, further confirming that naive, uninformed decision-making is inadequate for optimizing throughput under energy and access constraints.

\begin{figure}
    \centering
    \includegraphics[width=1\linewidth]{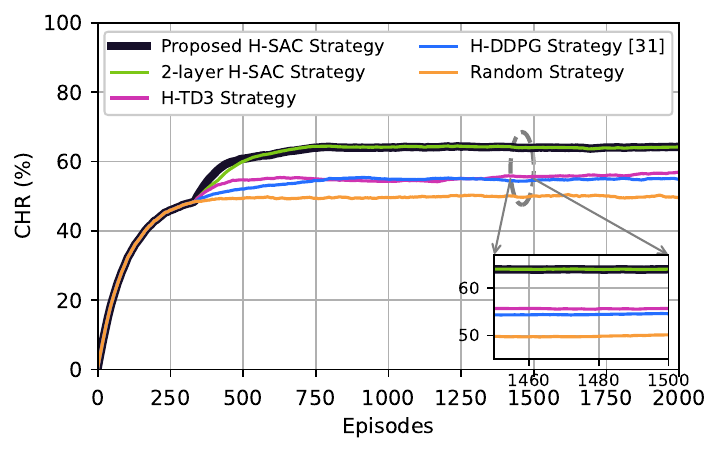}
    \caption{Benchmarking the CIoT hit rate of our proposed 
    H-SAC strategy in comparison to other existing strategies.}
    \label{fig:Benchmarking04}
\end{figure}

Fig.~\ref{fig:Benchmarking04} presents the performance of the CIoT cache hit rate over training episodes for the proposed H-SAC strategy compared to benchmark strategies. The H-SAC framework achieves the highest cache hit rate, converging near 65\% and maintaining consistent performance over time. This indicates the model’s ability to learn effective caching policies that anticipate user content requests while balancing energy and spectrum constraints. 
H-TD3 and H-DDPG achieve moderate hit rate performance, though both converge to noticeably lower values than H-SAC, and outperforming the random strategy. 
The 2-layer H-SAC strategy attains a CIoT hit rate comparable to that of the proposed H-SAC strategy, as both approaches are similar in design, with the distinction that 2-layer H-SAC employs a constant TS factor instead of learning it. Since the TS factor governs energy harvesting and transmission but does not influence caching decisions, their cache hit performance remains alike. However, in terms of ASR, the proposed H-SAC strategy substantially outperforms the 2-layer H-SAC strategy as highlighted in table~\ref{tab:comparison}. 
Overall, the results validate the importance of hierarchical and learning-based caching decisions in achieving high CIoT cache hit rates under constrained resources.

\begin{figure}
    \centering
    \includegraphics[width=1\linewidth]{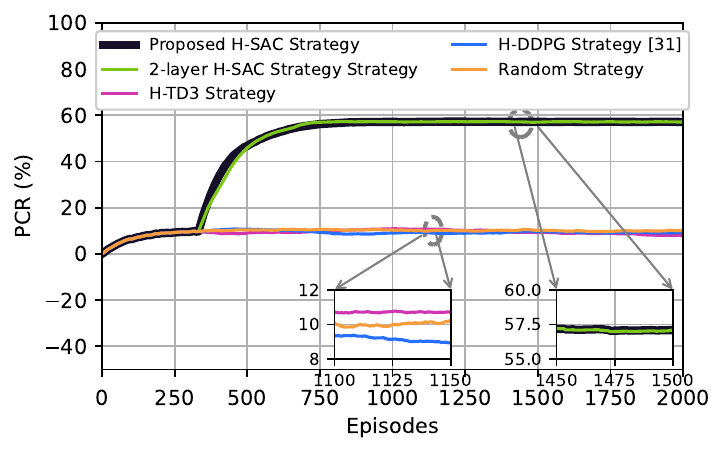}
    \caption{Benchmarking the cooperation rate of our proposed 
    H-SAC strategy in comparison to other existing strategies.}
    \label{fig:Benchmarking05}
\end{figure}

Fig.~\ref{fig:Benchmarking05} illustrates the cooperation rate between CIoT agents and the PU over training episodes for the proposed H-SAC strategy and the benchmarks. The cooperation rate in this context reflects how effectively SUs cache and serve PU-requested content, facilitating cooperative spectrum access. The proposed H-SAC strategy achieves the highest cooperation rate, converging above 55\%. This result demonstrates H-SAC’s ability to learn when and how to cache PU content strategically, using it as a bargaining mechanism for gaining spectrum access while adhering to energy and spectrum constraints.
Both H-TD3 and H-DDPG achieve cooperation rates comparable to the random strategy. However, their performance lags behind the proposed H-SAC approach. These results highlight the effectiveness of the H-SAC strategy in learning cooperative behavior with the PU for spectrum access, while H-TD3 and H-DDPG fail to develop such cooperation and instead behave in a largely random manner.
The 2-layer H-SAC strategy achieves a cooperation rate comparable to the proposed H-SAC strategy, similar to the CIoT hit rate in Fig.~\ref{fig:Benchmarking04}. 
Although the cache hit rate is similar, the proposed H-SAC strategy outperforms the 2-layer H-SAC strategy in terms of ASR, making it the preferable choice for the considered multi-objective problem as highlighted in table~\ref{tab:comparison}.

\begin{figure}
    \centering
    \includegraphics[width=1\linewidth]{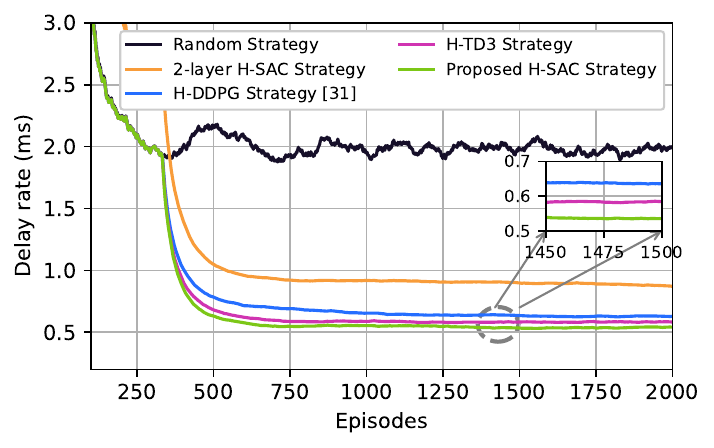}
    \caption{Benchmarking the delay of our proposed 
    H-SAC strategy in comparison to other existing strategies.}
    \label{fig:Benchmarking_delay}
\end{figure}

Fig.~\ref{fig:Benchmarking_delay} presents the delay performance over training episodes for the proposed H-SAC strategy and the benchmarking strategies. At the beginning of training, all agents behave randomly in order to fill their replay buffers. Once the buffers are sufficiently filled (around episode 333), the actual training process begins. At convergence, the proposed H-SAC strategy achieves the lowest delay among all strategies, while the Random strategy results in the highest delay, confirming the importance of adopting a learning-based approach. The 2-layer H-SAC strategy performs better than Random but worse than the other strategies, highlighting the importance of learning the time-switching factor $\rho$, which governs the balance between energy harvesting and transmission and thus directly impacts delay performance. Meanwhile, the H-TD3 and H-DDPG strategies outperform both Random and the 2-layer H-SAC in terms of delay but still fall short of the proposed H-SAC strategy. Taken together with the previous results, these findings demonstrate that our proposed H-SAC strategy is the most effective for achieving the optimal performance across all objectives in the considered multi-objective problem.

\begin{figure}
    \centering
    \includegraphics[width=1\linewidth]{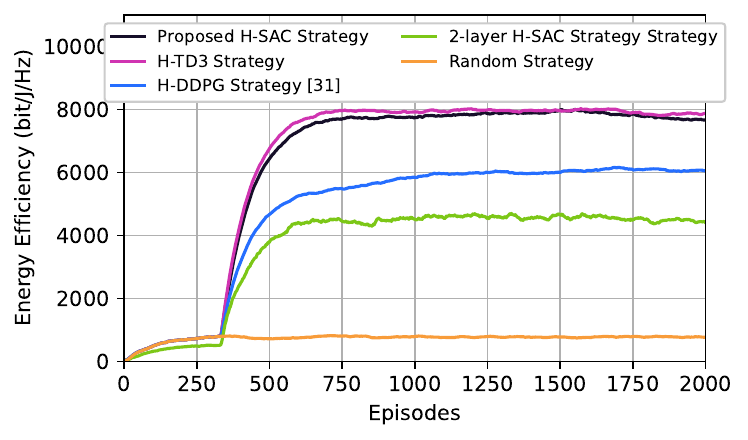}
    \caption{Benchmarking the EE of our proposed 
    H-SAC strategy in comparison to other existing strategies.}
    \label{fig:Benchmarking03}
\end{figure}

Fig.~\ref{fig:Benchmarking03} shows the performance in terms of EE over training episodes for the proposed H-SAC strategy and the benchmarking approaches. In terms of EE, the H-TD3 strategy slightly outperforms the proposed H-SAC. However, the difference is marginal when compared to H-SAC’s clear superiority in ASR, CIoT hit-rate, and cooperation rate.
 Thus, the H-TD3’s slight advantage in EE does not make it the better strategy, especially since the multi-objective problem was designed to prioritize maximizing ASR and hit rates while minimizing delay, as highlighted in table~\ref{tab:comparison}, with EE not being a primary objective. In contrast, H-DDPG exhibits significantly lower EE compared to both H-SAC and H-TD3. This indicates that H-DDPG is not only suboptimal for maximizing throughput but also incapable of effectively managing EH in energy-constrained devices.
The 2-layer H-SAC strategy results in consistently low EE, as it lacks dynamic adaptation to changing network and battery conditions. This highlights the drawbacks of static policy configurations in environments where EH, interference, and spectrum availability fluctuate. As expected, the random strategy performs the worst, offering no meaningful learning or adaptation over time and reinforcing the importance of intelligent, reinforcement learning-based optimization in CIoT systems.

\begin{figure}
    \centering
    \includegraphics[width=0.88\linewidth]{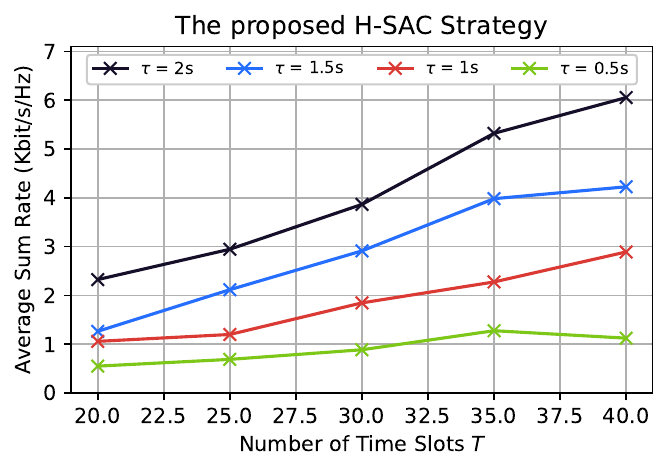}
    \caption{Illustrating the effect of varying the number of time slots $T$ and the time slot duration $\tau$ on the ASR performance of our proposed H-SAC strategy.}
    \label{fig:thrpts_T_tau}
\end{figure}

Fig.~\ref{fig:thrpts_T_tau} shows the effect of varying the number of time slots~\(T\) and their duration~\(\tau\) on the ASR performance of the proposed H-SAC strategy. As observed, increasing~\(T\) leads to a higher ASR. This can be attributed to the greater number of idle slots in which the PU is inactive, thereby reducing spectrum access restrictions. In such cases, the CIoT agent has more opportunities to transmit without interference constraints, allowing for more efficient utilization of harvested energy. Moreover, a larger~\(T\) increases the probability of favorable transmission conditions over the scheduling horizon. Similarly, increasing the duration~\(\tau\) of each slot improves ASR performance. Longer slots provide extended periods for both energy harvesting and data transmission, enabling the agent to accumulate sufficient energy before initiating high-throughput transmissions. This improvement is particularly beneficial under energy-constrained scenarios, where short slots may limit the agent's ability to both harvest and transmit effectively. It is also worth noting that the effects of~\(T\) and~\(\tau\) are complementary: a higher number of longer slots not only increases the frequency of transmission opportunities but also the quality of each transmission opportunity. However, in practical deployments, increasing~\(T\) or~\(\tau\) may be subject to latency requirements and network-level trade-offs, which should
\begin{figure}
    \centering
    \includegraphics[width=0.88\linewidth]{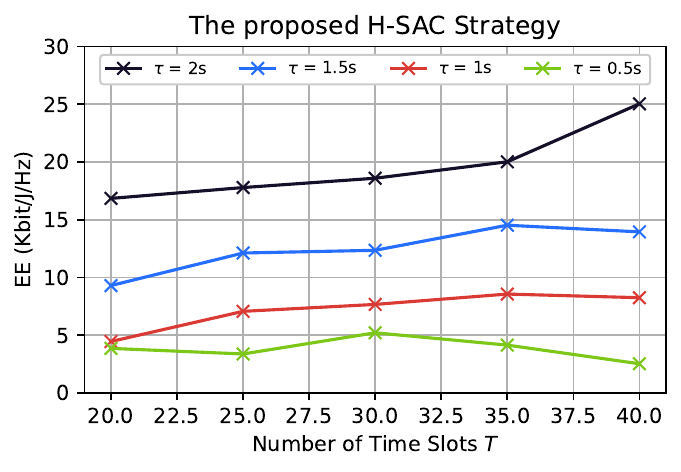}
    \caption{Illustrating the effect of varying the number of time slots $T$ and the time slot duration $\tau$ on the EE performance of our proposed H-SAC strategy.}
    \label{fig:ee_T_tau}
\end{figure}

Fig.~\ref{fig:ee_T_tau} illustrates the impact of varying the number of time slots~\(T\) and their duration~\(\tau\) on the EE performance of the proposed H-SAC strategy. Overall, increasing~\(T\) generally leads to higher EE, except when \(\tau = 0.5\)~s. This deviation can be attributed to the limited time available for energy harvesting (EH) in each slot when \(\tau\) is small, resulting in unstable energy consumption patterns. In contrast, for larger~\(\tau\) values, the trend is more consistent, and EE steadily improves with increasing~\(T\), as longer slots allow for more efficient EH and energy utilization.
It is worth noting that, for \(\tau = 0.5\)~s, the ASR trend still increases with~\(T\), whereas the EE trend does not follow the same pattern. This indicates that the drop in EE is primarily due to suboptimal EH decisions and reduced opportunities for energy accumulation, rather than limitations in throughput. This behavior can also be linked to the multi-objective optimization goal, which prioritizes maximizing throughput and cache hit rates while minimizing delay, without explicitly optimizing for EE. As a result, the policy may sacrifice energy efficiency in favor of the other performance objectives.

\begin{figure}
    \centering
    \includegraphics[width=0.9\linewidth]{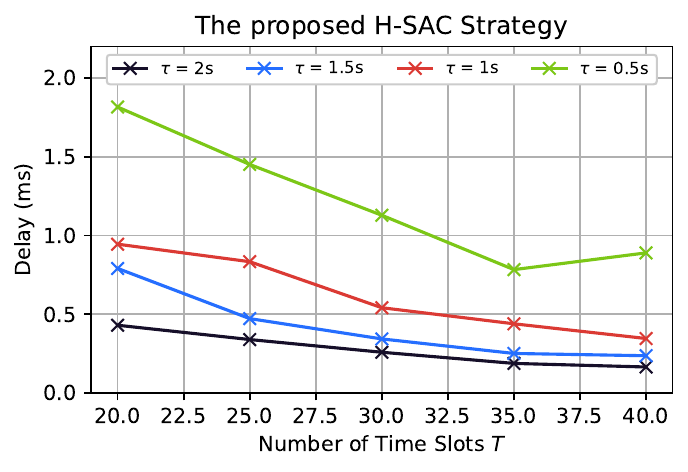}
    \caption{Illustrating the effect of varying the number of time slots $T$ and the time slot duration $\tau$ on the delay performance of our proposed H-SAC strategy.}
    \label{fig:dly_T_tau}
\end{figure}

Fig.~\ref{fig:dly_T_tau} presents the effect of varying the number of time slots~\(T\) and their duration~\(\tau\) on the delay performance of the proposed H-SAC strategy. The results show that increasing~\(T\) leads to a steady reduction in delay, while increasing~\(\tau\) results in a substantial drop in delay across all values of~\(T\). For example, when \(\tau\) increases from \(0.5\,\text{s}\) to \(1.5\,\text{s}\), the delay decreases by up to~\(45\%\) for \(T=20\). This improvement can be attributed to the fact that a larger~\(T\) provides the agent with more idle time slots during which the PU is inactive, allowing transmission at higher power levels without interference constraints. Consequently, higher transmission rates reduce the overall delay. Similarly, increasing~\(\tau\) extends the portion of each time slot dedicated to EH, enabling the agent to accumulate more energy in its battery and subsequently transmit at higher powers, further reducing delay.

\begin{figure}
    \centering
    \includegraphics[width=0.9\linewidth]{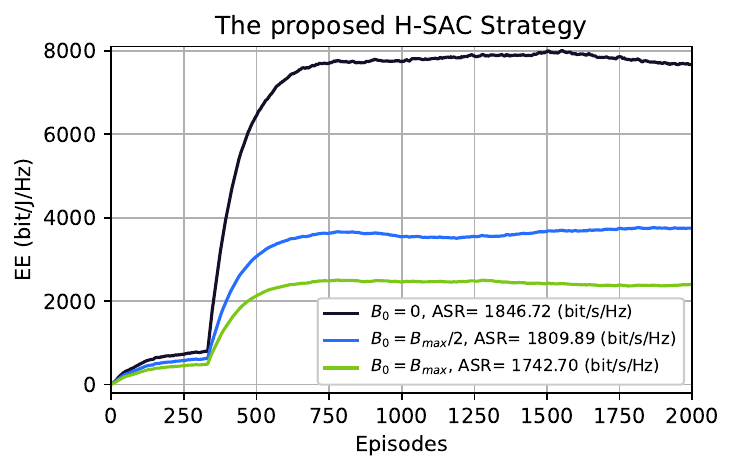}
    \caption{Illustrating the effect of varying the initial battery level $B_0$ on the EE performance of our proposed H-SAC strategy.}
    \label{fig:ee_B0}
\end{figure}

In Fig.~\ref{fig:ee_B0}, we illustrate the impact of varying the initial battery level~\(B_0\) on the EE performance of the proposed H-SAC strategy. As observed, starting with a full battery results in the lowest EE. This suggests that a high initial charge may hinder the agent’s ability to learn an optimal EH policy, ultimately reducing energy efficiency. Conversely, when the battery is initially empty, the agent achieves significantly higher EE. This improvement may be attributed to the fact that with a low initial charge, the agent experiences penalties in the early time slots, encouraging exploration of alternative actions and leading to a more effective EH strategy. Moreover, examining the ASR values reveal that \(B_0 = 0\) achieves an ASR of \(1846.72~\text{bits/s/Hz}\), slightly higher than the other two \(B_0\) values, while the EE converges to around \(8\times10^3\), compared to less than \(4\times10^3\) for the others. This indicates that the improved EE performance is not primarily driven by transmission ASR, but rather by a more efficient EH and energy consumption policy learned by the agent.

\begin{table}
\centering
\caption{Comparison of performance parameters at convergence for different strategies.}
\label{tab:comparison}
\resizebox{\columnwidth}{!}{%
\begin{tabular}{|c|c|c|c|c|c|c|}
\hline
\diagbox[width=8em]{\textbf{Strategy}}{\textbf{Parameter}} 
& \textbf{Reward} 
& \makecell{\textbf{ASR}\\ \textbf{(bits/s/Hz)}} 
& \textbf{CHR}
& \textbf{PCR} 
& \makecell{\textbf{Delay}\\ \textbf{(ms)}}
& \makecell{\textbf{EE}\\ \textbf{(bits/J/Hz)}}\\
\hline
\makecell{\rule{0pt}{0.1ex}\\ \textbf{Proposed}\\ \textbf{H-SAC}\\ \rule{0pt}{0.1ex}} & 713 & 1846.72 & 64.2\% & 57.1\% & 0.54 & 7670.16\\
\hline
\makecell{\rule{0pt}{0.1ex}\\\textbf{H-TD3}\\ \rule{0pt}{0.1ex}}   & 659 & 1716.50 & 56.9\% & 7.8\% &0.58 & 7878.56\\
\hline
\makecell{\rule{0pt}{0.1ex}\\ \textbf{H-DDPG}\\ \rule{0pt}{1ex}}  & 602 & 1589.06  & 54.8\% & 9.3\% &0.62 & 6052.85\\
\hline
\makecell{\rule{0pt}{0.1ex}\\ \textbf{2-layer H-SAC}\\ \rule{0pt}{1ex}} & 449 & 1147.34  & 64.1\% & 57.1\% & 0.87& 4414.28\\
\hline
\makecell{\rule{0pt}{0.1ex}\\ \textbf{Random}\\ \rule{0pt}{1ex}}  & -55 & 501.40  & 49.6\% & 10.1\% &1.99& 771.25\\
\hline
\end{tabular}
}
\end{table}
Table~\ref{tab:comparison} provides a comprehensive comparison of the proposed H-SAC framework against baseline strategies, displaying their performance at convergence across reward, ASR, CHR, PCR, delay, and EE. The results in Table~\ref{tab:comparison} demonstrate that the proposed H-SAC strategy consistently outperforms all other approaches across the key performance metrics. In particular, H-SAC achieves the highest reward and ASR, while also maintaining superior PCR and CHR. The only exception is in EE, where H-TD3 shows a slightly higher value; however, EE is not a primary objective in the considered multi-objective problem, making this trade-off acceptable. It is also observed that the CHR and PCR of H-SAC are similar to those of the 2-layer H-SAC strategy. This behavior is expected, as both strategies share identical caching and cooperation mechanisms, with the main distinction being that 2-layer H-SAC employs a constant TS factor rather than learning it. Since the TS factor directly influences EH and transmission decisions, the 2-layer H-SAC strategy suffers significantly in terms of reward, ASR, and EE. This comparison underscores the importance of learning all parameters, while fixing one parameter may yield comparable caching outcomes, it severely limits overall system performance. The proposed H-SAC strategy achieves the lowest delay (0.54 ms), showing its efficiency in minimizing service latency. H-TD3 (0.58 ms) and H-DDPG (0.62 ms) perform reasonably well but remain less effective than H-SAC. 2-layer H-SAC incurs higher delay (0.87 ms) due to its lack of adaptability, while random performs worst (1.99 ms). These results confirm that adaptive parameter learning is crucial for reducing delay in latency-sensitive CIoT environments. It is worth noting that both ASR and EE are expressed per unit bandwidth, and in this study, we normalize the bandwidth in order to generalize the results. In real-world deployments, where the available bandwidth is significantly larger, the absolute performance differences between strategies would scale accordingly, making the superiority of H-SAC even more pronounced. 

\section{Conclusions}
In this work, we proposed a novel hierarchical DRL framework based on the SAC and DQN algorithms to address joint energy management, hybrid spectrum access, power control, and cooperative caching in CIoT networks. By decomposing the decision-making process into three levels, TS control, access coordination (cache sharing), and low-level power and caching actions, our approach effectively manages the complexity of mixed discrete-continuous action spaces and adapts to dynamic wireless environments. 
\textcolor{black}{Adaptive learning of the TS factor is pivotal. Fixing it degrades throughput and energy efficiency despite comparable cache performance, underscoring the necessity of dynamic EH control. Finally, cooperative caching provides an effective mechanism for spectrum access, increasing PU cooperation without compromising the CIoT hit rate when coupled with interference-aware power control.} 
Extensive simulations under channel fading, uncertain content demand, and energy limitations demonstrate that the proposed H-SAC framework improves average sum rate, reduces latency, and enhances cache efficiency and energy utilization.

\textcolor{black}{Despite these gains, several challenges remain for real-world deployment, including the computational overhead of training hierarchical policies on resource-constrained devices, the need for sample-efficient online learning under partial observability, and the communication overhead associated with cooperative caching. Future work will investigate adaptive weight tuning, non-linear EH models, heterogeneous file sizes, and testbed-based evaluations to bridge the gap between analysis and practice. We also plan to extend the framework to collaborative multi-agent settings, where coordinated policies can further improve spectrum efficiency and cache diversity.}

\bibliography{ref.bib}
\bibliographystyle{IEEEtran}
\begin{IEEEbiography}[{\includegraphics[width=1in,height=1.25in,clip,keepaspectratio]{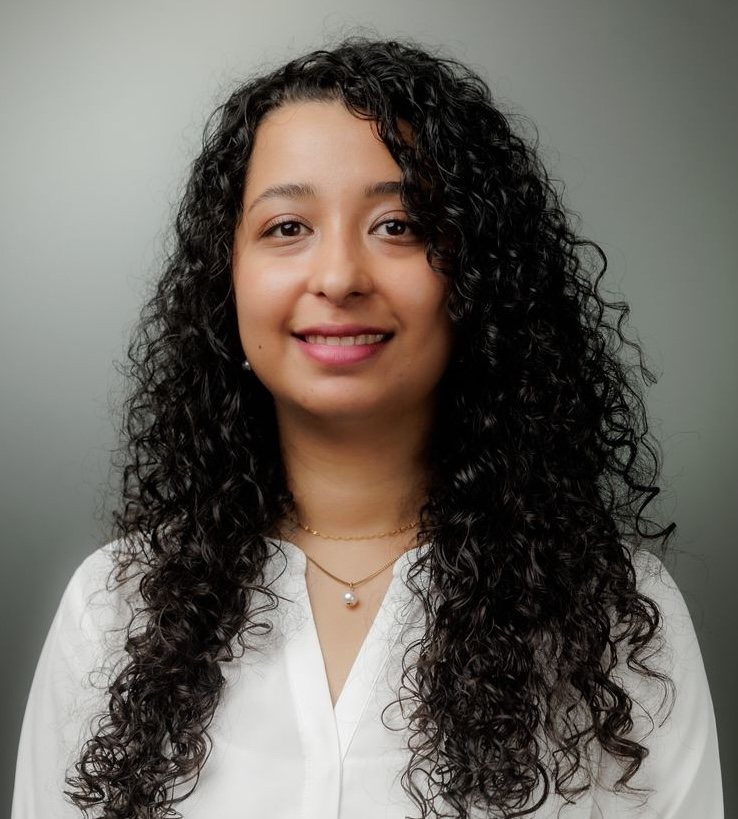}}]{Nadia Abdolkhani} 
(Student Member, IEEE) received the B.Sc. and M.Sc. degrees in electrical engineering from Shiraz University of Technology, Shiraz, Iran, in 2017 and 2021, respectively. Currently, she is pursuing a Ph.D. from the Department of Electrical and Computer Engineering at Concordia University, Montreal, Canada. Her research interests include wireless and cellular communications, caching at wireless networks, device-to-device (D2D) communications, intelligent wireless communications, including deep learning and reinforcement learning-driven network operations, cognitive and cooperative communications. She is a recipient of Concordia University's International Tuition Award of Excellence.
\end{IEEEbiography}

\begin{IEEEbiography}[{\includegraphics[width=1in,height=1.25in,clip,keepaspectratio]{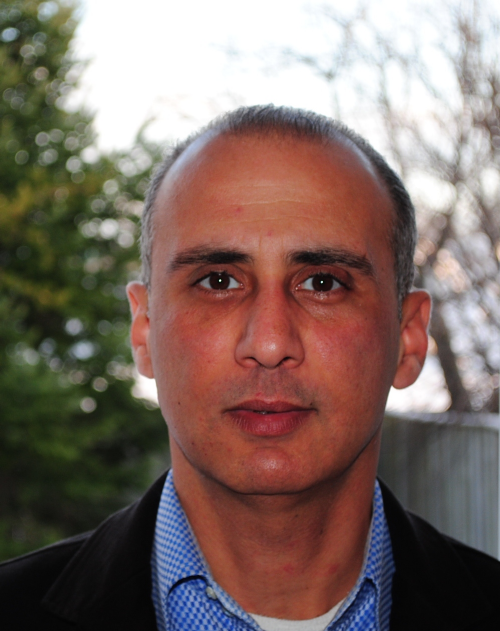}}]{Walaa Hamouda} (Senior Member, IEEE) received the M.A.Sc. and Ph.D. degrees in electrical and computer engineering from Queen’s University, Kingston, ON, Canada, in 1998 and 2002, respectively. Since July 2002, he has been with the Department of Electrical and Computer Engineering, Concordia University, Montreal, QC, Canada, where he is currently a Professor. Since June 2006, he is Concordia University Research Chair Tier I in Wireless Communications and Networking. His current research interests include machine-to-machine communications, IoT, 5G and beyond technologies, single/multiuser multiple-input multiple-output communications, space-time processing, cognitive radios, wireless networks. Dr. Hamouda served(ing) as Co-chair of the IoT and Sensor Networks Symposium of the GC’22, TPC Co-chair of the ITC/ADC 2022 conference, track Co-Chair: Antenna Systems, Propagation, and RF Design, IEEE Vehicular Technology Conference (VTC Fall'20), Tutorial Chair of IEEE Canadian Conference in Electrical and Computer Engineering (CCECE 2020), General Co-Chair, IEEE SmartNets 2019 Conference, Co-Chair of the MAC and Cross Layer Design Track of the IEEE (WCNC) 2019,  Co-chair of the Wireless Communications Symposium of the IEEE ICC’18, Co-chair of the Ad-hoc, Sensor, and Mesh Networking Symposium of the IEEE GC’17, Technical Co-chair of the Fifth International Conference on Selected Topics in Mobile \& Wireless Networking (MoWNet’2016), Track Co-Chair: Multiple Antenna and Cooperative Communications, IEEE Vehicular Technology Conference (VTC Fall'16), Co-Chair: ACM Performance Evaluation of Wireless Ad Hoc, Sensor, and Ubiquitous Networks (ACMPE-WASUN'14) 2014, Technical Co-chair of the Wireless Networks Symposium, 2012 Global Communications Conference, the Ad hoc, Sensor, and Mesh Networking Symposium of the 2010 ICC, and the 25th Queen’s Biennial Symposium on Communications. He also served as the Track Co-chair of the Radio Access Techniques of the 2006 IEEE VTC Fall 2006 and the Transmission Techniques of the IEEE VTC-Fall 2012. From September 2005 to November 2008, he was the Chair of the IEEE Montreal Chapter in Communications and Information Theory. He is an IEEE ComSoc Distingushed Lecturer. He received numerous awards, including the Best Paper Awards of the IEEE GC 2023, IEEE GC 2020, IEEE ICC 2021, ICCSPA 2019, IEEE WCNC 2016, IEEE ICC 2009, and the IEEE Canada Certificate of Appreciation in 2007 and 2008. He served as an Associate Editor for the IEEE COMMUNICATIONS LETTERS, IEEE TRANSACTIONS ON SIGNAL PROCESSING, IEEE COMMUNICATIONS SURVEYS AND TUTORIALS, IET WIRELESS SENSOR SYSTEMS, IEEE WIRELESS COMMUNICATIONS LETTERS, TRANSACTIONS ON VEHICULAR TECHNOLOGY, and currently serves as an Editor for the IEEE TRANSACTIONS ON COMMUNICATIONS, IEEE TRANSACTIONS ON WIRELESS COMMUNICATIONS, IEEE IoT journal.
\end{IEEEbiography}

\end{document}